# A prototype optical link board with redundancy design for the ATLAS liquid argon calorimeter Phase-2 upgrade

X. Huang,[a,b] C. Liu,[a] A. Deiana,[a] B. Deng,[c] D. Gong,[a] L. Hasley,[a] S. Hou,[d] T. Liu,[c,*] H. Sun,[a,b,1] J. Ye,[a] L. Zhang,[a,b,1] and W. Zhang[a,b,1]

[a] *Southern Methodist University,*
*Dallas, TX 75275, USA*

[b] *Central China Normal University,*
*Wuhan, Hubei 430079, China*

[c] *Hubei Polytechnic University,*
*Huangshi, Hubei 435003, China*

[d] *Academia Sinica,*
*Nangang, Taipei 11529, Taiwan*

E-mail: tliu@smu.edu

ABSTRACT: A prototype optical-link board has been developed for the ATLAS Liquid Argon Calorimeter Phase-2 upgrade. The board consists of 24 lpGBT chips and 8 VTRx+ modules and demonstrates the full optical link design of the future front-end board. The board has 22 simplex optical links to transmit detector data, which are emulated in FPGAs and injected through 6 FMC connectors, to the off-detector electronics. The board implements 2 duplex optical links for clocks, control, and monitoring with redundant design to improve the system reliability. Data transmission and all redundant designs have been verified. A fiber routing scheme, in which 2 or 3 VTRx+ modules are re-connected to the front panel with an MPO connector, has been prototyped.

---

[*] Corresponding author.
[1] Visiting scholars at SMU and performed this work at SMU.

# Contents

## 1. Introduction

The readout electronics of the ATLAS Liquid Argon Calorimeter is under development for High-Luminosity Large Hadron Collider (HL-LHC) [1]. For the front-end readout electronics, the upgrade aims to provide full detector granularity to the trigger system in the so-called free-running scheme by sending all data off detector. After the upgrade, the data rate of each Front-End Board (FEB), including 128 detector channels, will be higher than 200 Gb/s. The total data rate of the whole LAr calorimeter, including 1524 FEBs, is higher than 300 Tb/s. The upgrade also ensures adequate radiation tolerance. The front-end electronics operates in a harsh radiation environment with a Total Ionizing Dose (TID) of 2.25 kGy, a Non-Ionizing Energy Loss (NIEL) fluence of 4.9 x $10^{13}$ $n_{eq}$/cm$^2$, and a SEE fluence of 7.7 x $10^{12}$ h/cm$^2$.

    All detector data are transmitted through optical links. To verify the design of the optical link part on the future FEB, based on the simplex data link demonstration board [2], we have designed a prototype-link board. Figure 1 is the block diagram of the FEB with the optical link part marked in a yellow box. Each FEB is divided into two halves. For each half FEB, 64 detector channels are amplified and shaped in 16 4-channel Pre-Amplifiers (PAs, including shapers) [3] and digitized in 16 8-channel Analog-to-Digital Converters (ADCs) [4]. Each detector channel is digitized with two different gains (H and L shown in the figure standing for high and low, respectively) simultaneously to achieve a 16-bit dynamic range with 12-bit-resolution ADCs. The digitized data are transmitted off detector through upstream optical links called data links. We also need control links to provide clocks, Bunch Crossing Reset (BCR) signals, I$^2$C configuration, and other slow control/monitoring (such as resets and temperatures) support. Each FEB has 22 simplex data links and 2 duplex control links. Optical links are based on lpGBTs [5] and VTRx+ modules [6]. Each FEB has 24 lpGBTs and 8 VTRx+ modules. About forty thousand lpGBTs and thirteen thousand VTRx+ modules will be deployed in the calorimeter front-end electronics. The lpGBTs responsible for data links will be called data lpGBTs, and the lpGBTs responsible for control links will be called control lpGBTs.

    If a data link is broken, we lose the data from up to 6 detector channels, whereas if a control link is broken, we lose control of half FEB or 64 detector channels. To improve reliability, we implement redundant control links. Based on past experiences [7-8], Integrated Circuits (ICs) are more reliable than fibers and optical modules. In the redundancy design, we assume that lpGBTs are still operational even if an optical module or a fiber that the lpGBT is

connected to is broken down. The redundancy requires some communication between the two halves, marked a bidirectional arrow between two halves in the figure.

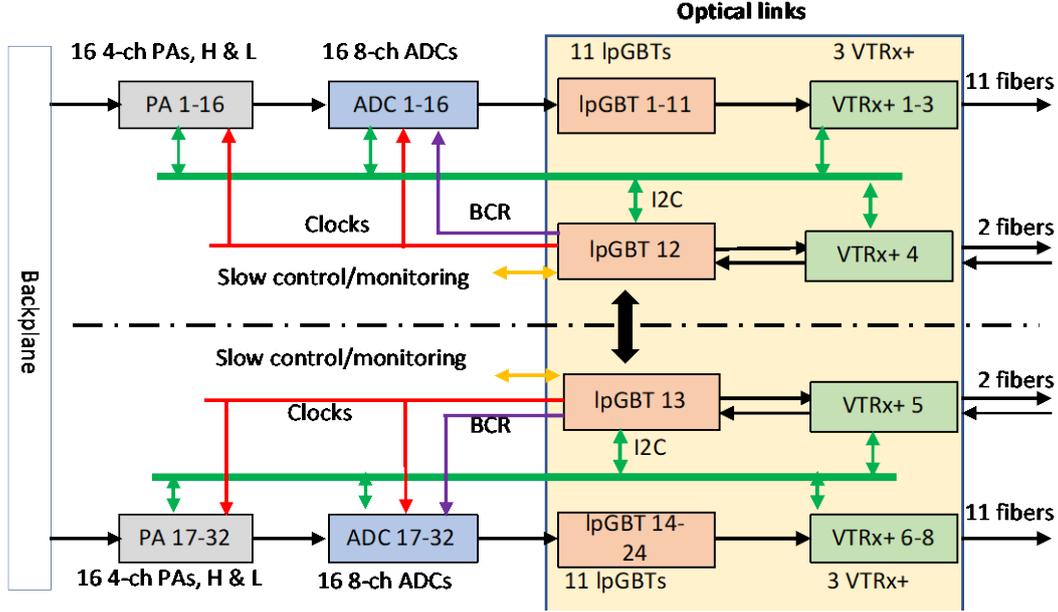

Figure 1. Block diagram of the FEB.

## 2. Design of the prototype link board and verification of data transmission

The prototype-link board implements the full optical link part of the future FEB. The 22 lpGBT chips operating in the transmitter mode and 6 VTRx+ modules with only transmitter channels enabled, and 22 simplex optical fibers form data links. Each data link transmits detector data at 10.24 Gb/s. The data are emulated in FPGAs and injected into the board through 6 FPGA Mezzanine Card (FMC) connectors. Two lpGBT chips, two VTRx+ modules, and two pairs of duplex fibers constitute the control links. These lpGBT chips operate in the transceiver mode and each VTRx+ module has a transmitter channel and the receiver channel enabled. Each downstream control link operates at 2.56 Gb/s, and each upstream control link operates at 5.12 Gb/s. DC-DC converts and Low-Drop-Out (LDO) voltage regulators are included to generate all voltages for lpGBTs and VTRx+ modules. Voltages, currents, and temperatures at ten positions are monitored through control links. The Prototype-link board size is 490 mm x 134 mm x 2 mm. The board has 14 layers and adopts the same layer stack and low-loss halogen-free material (EM-888K) as the future FEB2 so that the design (including layout) can be easily integrated into the future FEB2. Figure 2 is the photograph of the prototype-link board and the test setup.

    Each VTRx+ module has a fiber pigtail with a Mechanical Transfer (MT) connector. We will have to connect 8 VTRx+ modules with three MPO adapters mounted on the front panel. The routing from 3 VTRx+ modules to the front panel with a 3-MT-to-1-MPO fiber splitter is shown at the right of figure 2. The connection from VTRx+ modules to the front panel costs extra fibers, insertion loss (MT-to-MT connection), board area (MT-MT clamps), and assembly effects. To overcome these routing issues, we are exploring the possibility of re-connecting two or three VTRx+ modules directly to an MPO adapter. The prototype samples with such re-connection fibers are shown at the left of figure 2. The current prototypes have fibers longer than needed for possible re-working. In the future, the fiber length can be optimized after the

board layout is finalized. Note that in figure 2, we added an extension board between the prototype-link board and the front panel. The area of the extension board is reserved for DC-DC converts and LDO regulators for PAs and ADCs.

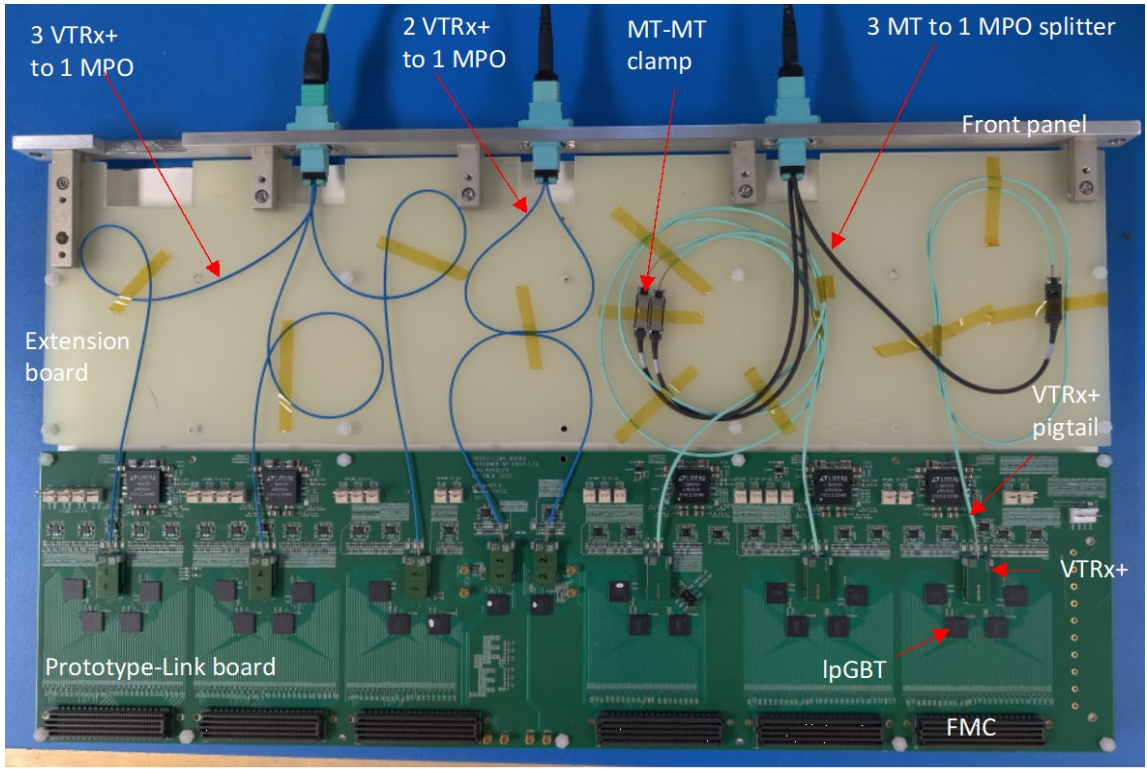

Figure 2. Photograph of the prototype-link board.

During the test, an Si5338 Evaluation board provides clocks for the whole test setup. Three FPGA evaluation boards (Xilinx KC705) are used, one to emulate the detector data and two to emulate the off-detector electronics. We have verified the data transmission of the prototype-link board. Typical optical eye diagrams of all upstream and downstream links are shown in figure 3. The Bit Error Rates (BER) of all data links were measured. For each link, no error occurred for at least 3 minutes, corresponding to a BER of $10^{-12}$ at 10.24 Gb/s with a confidence level of 84%.

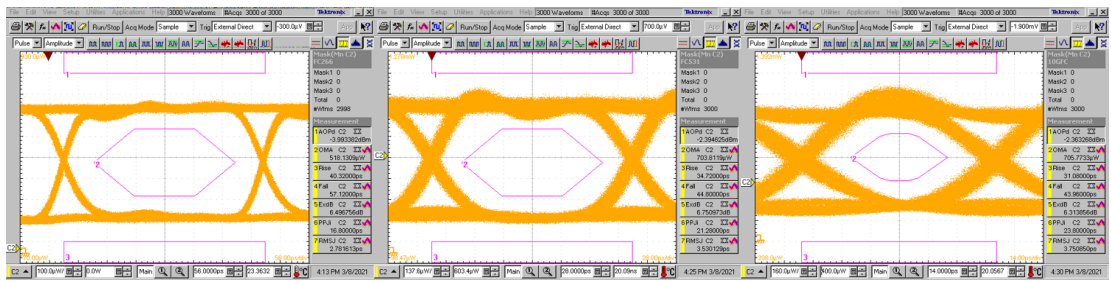

Figure 3. Eye diagrams of a downstream control link at 2.56 Gb/s (left), the upstream control link at 5.12 Gb/s (middle), and an upstream data link at 10.24 Gb/s (right).

## 3. Design and verification of redundant control links

Figure 4 is the redundant design of the clock distribution. Each PA needs a fixed-phase clock at 40 MHz. Each ADC needs two phase-adjustable clocks, one at 40 MHz and the other one at 640 MHz. Each control lpGBT (12 and 13) provides 40-MHz clocks to the data lpGBTs and these data lpGBT chips generate clocks for PAs and ADCs on the same half FEB board. If VTRx+ 5 or its downstream fiber is broken, lpGBT 12 supplies a 40-MHz clock to lpGBT 13 (Pins RefClk) and controls configuration pins (Mode1 and Lock Mode) of lpGBT 13 via general-purpose I/O pins of lpGBT 12. We have verified in the test that we can generate all clocks with a downstream control link unplugged.

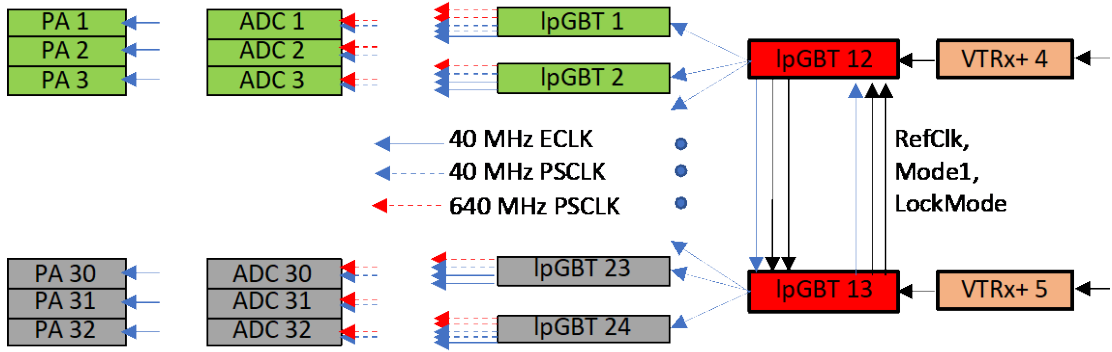

**Figure** 4. Redundant design of the clock distribution.

Each ADC needs a BCR signal to reset its internal Bunch Crossing IDentification (BCID) counter for data alignment on the off-detector electronics. The BCR signal has a pulse lasting for a bunch crossing cycle (25 ns) in a whole period of 3564 bunch crossing cycles. The control links provide 32 BCR signals to 32 ADCs through 32 e-Tx ports [9] of two control lpGBTs. The baseline redundancy will be implemented in the new ADC [10], which has been submitted, though an alternative redundant design has been implemented and verified on the prototype link board. In the baseline redundant design, each new ADC has two BCR inputs, but only one BCR input is active. Each BCR signal is multiple dropped to two ADCs. The termination of the first ADC is turned off, whereas the termination of the second ADC is turned on. We can select which BCR input pin is active through the $I^2C$ configuration. The baseline design will be verified in the future.

Figure 5 is the redundant design of the $I^2C$ configuration. Though all ASICs (PAs, ADCs, lpGBTs, VTRx+) have default register values, they can be remotely configured. Especially, each data VTRx+ module, which has a transmitter channel and a receiver channel enabled in default, has to be configured to operate with only transmitter channels enabled after each power cycle. In the normal $I^2C$ operation, lpGBTs 12 and 13 are configured through their Internal Control (IC) channels, while lpGBTs 11 and 14 are configured via their External Control (EC) channels. Each lpGBT has three $I^2C$ controllers (M0, M1, and M2). VTRx+ 3 and 6 are configured through M2 of lpGBTs 11 and 14. The rest lpGBTs, VTRx+, PAs, and ADCs are configured through M0-M2 of lpGBTs 12 and 13. Note that two controllers (a primary one and a secondary one) connect to an $I^2C$ bus. The control lpGBTs 12 and 13 are the primary $I^2C$ controllers. In case of VTRx+ 4 or its fiber is broken, lpGBT 12 controls the SC_I2C pin of lpGBT 13 and 14 to switch their configuration mode, and lpGBTs 10 and 11, as the secondary controllers, take over the I$^2$C buses, which are originally controlled by M0-M2 of lpGBT 12. We have verified that we can configure the whole board through one duplex control link.

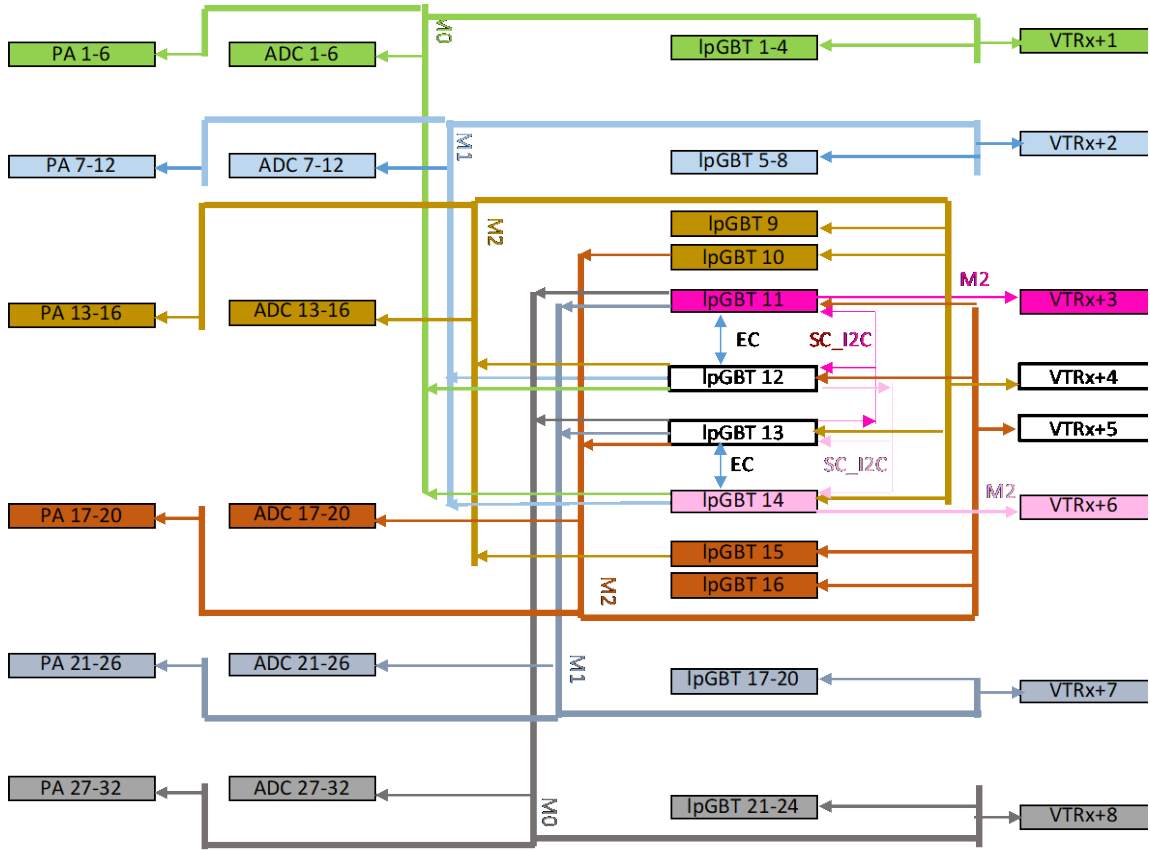

Figure 5. Redundant design of the I$^2$C configuration.

In special cases, an ASIC may be stuck and need to be reset or power cycled. All ASICs on the FEB can be reset or power cycled. Particularly, lpGBT 12 can reset or power cycle lpGBT 13 and vice versa. The crossover reset and power cycle design has been verified on the prototype-link board.

**4. Conclusion**

A prototype optical-link board has been developed for the ATLAS Liquid Argon Calorimeter Phase-2 upgrade. The board consists of 24 lpGBT chips and 8 VTRx+ modules and demonstrates the full optical link design. The board has 22 simplex optical links to transmit detector data to the off-detector electronics. Full data transmission has been verified. The board implements 2 duplex optical links for clocks, control, and monitoring with a redundant design to improve the system reliability. All redundant designs have been verified. A fiber routing scheme, in which 2 or 3 VTRx+ modules are re-connected to the front panel with an MPO connector, has been prototyped. The functionality and hardware implementation of the prototype board can be integrated into the future FEB2.

## Acknowledgments

We acknowledge the support of the US-ATLAS R&D grant for the ATLAS LAr Phase-II upgrade project.